\documentclass[onecolumn,showpacs,floatfix,11pt,nofootinbib]{revtex4}
\usepackage{graphicx}
\usepackage{longtable}

\RequirePackage{amssymb}

\begin{document}

\title{On the Thermodynamical Black Hole Stability in the Space-time of a Global Monopole in $f(R)$-Gravity}

\author{F. B. Lustosa$^{1}$}
\email{chicolustosa@gmail.com}
\author{M.E.X.Guimar\~aes$^{2}$}
\email{emilia@if.uff.br}
\author{Cristine N. Ferreira$^{3}$}
\email{crisnfer@pq.cnpq.br}
\author{J. L. Neto$^{4}$}
\email{jlneto@if.ufrj.br}
\author{J. A. Helayel-Neto$^{5}$}
\email{helayel@cbpf.br}

\affiliation{1. Instituto de F\'{\i}sica, Departamento de F\'{\i}sica Te\'orica, Universidade do Estado do Rio de Janeiro, Rua S\~ao Francisco Xavier, 524, 20550-013, Maracan\~a, Rio de Janeiro, RJ, Brazil}

\affiliation{2. Instituto de F\'{\i}sica, Universidade Federal
Fluminense, Av. Gal. Milton Tavares de Souza, s/n - Campus da
Praia Vermelha 24210-346 Niter\'oi, RJ, Brazil}

\affiliation{3. N\'ucleo de Estudos em F\'{\i}sica, Instituto Federal de Educa\c{c}\~ao Ci\^encia e Tecnologia
Fluminense, Rua Dr Siqueira, RJ, Brazil}

\affiliation{4. Instituto de F\'{\i}sica, Universidade Federal do Rio de Janeiro, Caixa Postal 68528 Rio de Janeiro 21941-972, RJ, Brazil}

\affiliation{5. Centro Brasileiro de Pesquisas F\'{\i}sicas, Rua Dr. Xavier Sigaud 150, Urca, Rio de Janeiro, Brazil}

\pacs{?}

\begin{abstract}

 In this work, we re-assess a class of black hole solutions in a global monopole spacetime in the framework of an  $f(R)$-gravity model. Our main line of investigation consists in considering a region close enough to the black hole,  but such that the weak field approximation is still valid. The stability of the black hole is studied in terms of its thermodynamical properties, with the radial coordinate written as a power law function with the status of the main factor underneath the stability of the model.  We obtain  the explicit expressions for the thermodynamical quantities of the black hole as functions of the event horizon, by considering both the Hawking and the local temperatures.  The phase transitions that may occur in this system,  including the Hawking-Page phase transition, are inspected with particular attention. We work out and contemplate a solution of special interest in which one of the parameters is related to the cosmological constant. Our main result sets out to establish a comparison between 
 both the Hawking and the local formalisms for the black hole in the framework of the $f(R)$-gravity in the particular space-time adopted here.

 \end{abstract}
\maketitle

\newpage

\section{Introduction}

There are still some important  issues for which current cosmological standard theories in Physics have not yet found complete answers.  General Relativity (GR) has not been able to clarify questions such as  the non-renormalization of the gravity theory, the singularity problems in black hole physics and the physics of the early Universe,  leading us to the necessity of finding alternative approaches. In this context, one of the most intriguing facts is the accelerating Universe. Despite of the existence of some alternative approaches  to explain this behavior, one of those without adding dark matter or dark energy \cite{Buchdahl:1983zz,Morais:2015ooa,Takahashi:2015ifa}, and the other with  $f(R)$ gravities \cite{Nojiri:2010wj,Sotiriou:2008rp} that has received much attention as one of the strongest candidates to explain the current accelerating universe \cite{Perlmutter:1998np}. The $f(R)$  gravity is constructed by replacing the Ricci scalar in the Einstein-Hilbert action by an arbitrary function of the Ricci scalar $f (R)$. Such theory is well known to lead to an extra scalar degree of freedom and it has been shown, for example in \cite{Hu:2007nk,Appleby:2007vb,Starobinsky:2007hu}, that observationally acceptable models may be built up. In some new contributions, f(R)-gravity can exhibit dependence on higher powers of the curvature scalar \cite{Ohta:2015efa}. Indeed, it must be stressed that the recent starting of the gravitational wave astronomy through the LIGO detections could be, in principle, an important tool to test the validity of the f(R) theories of gravity, as stated in the ref. \cite{corda}.

In 1972, Bekenstein published his first article demonstrating  the relation between thermodynamic quantities and gravitational properties of a black hole. His works \cite{Bekenstein:1972tm}-\cite{Bekenstein:1974ax} were followed by the model of particle creation around BH proposed by Hawking \cite{Hawking:1974sw}. Since then, the thermodynamical  behavior of quasi-classical systems has been widely explored in other gravitational frameworks. Most recent contributions were made in the study of the thermodynamics of the black hole in  modified Schwarszchild \cite{Cai:2009ua}, Born-Infeld-anti-de Sitter \cite{Myung:2008eb},  Horava-Lifshitz \cite{Biswas:2010zzb} and  $f(R)$ theory \cite{delaCruzDombriz:2009et}.

It has been shown that topological defects such as  cosmic strings, monopoles and domain walls
could be formed as a result of spontaneously broken symmetry in a vacuum phase transition of the early universe \cite{Kibble:1976sj}-\cite{Vilenkin:1984ib}. In the context of the GR, Barriola and Vilenkin \cite{Barriola:1989hx} studied the gravitational effects of a global monopole as a spherically symmetric topological defect. In the context of $f(R)$-theories, global monopoles heve been recently studied by \cite{Carames:2011uu}-\cite{Carames:2011xi}, and references therein. In this solution, there appears a term which corresponds to a black hole solution.
It is natural, then, to imagine that there is a region with a global monopole that could have been swallowed by a black-hole.  Another interesting situation where the monopole appears as an important ingredient involves boson stars. When these compact objects are present along with a non-minimally coupled global monopole, a black-hole-type solution can be mimicked. This approach is important to unveil the non-linear gravitational effects and the gravitational back reaction \cite{Marunovic:2013eka}.

On the other hand, the analysis of the thermodynamical properties of that case, in the context of GR, was carried out in \cite{Yu:1994fy} and, more recently, in the context of an $f(R)$-theory in \cite{Man:2013sf}, where  the authors adopt the weak field approximation solution presented in \cite{Carames:2011uu} with a specific {\it ansatz} for $ f'(R) = F(R) = \frac{df(R)}{dR} $ as a linear function of the radial coordinate.

In the present paper, we consider thermodynamical aspects of black holes in the space-time of a global monopole in the framework of $f(R$)-gravity.
We study a general case in which  $f(R)$ is a power law function of the radial coordinates. We anticipate that we obtain the explicit expressions for the local thermodynamical quantities of the black hole as a function of the event horizon, the parameter describing the monopole and the measurable corrections on the usual GR gravity due to the $f(R)$-extension.
The paper is organized as follows.
In Section 2, motivated by the analysis of thermodynamic aspects, we revise the solution for a black hole in the region containing a global monopole with the use of a $ f (R) $ theory in the weak field approximation \cite{Carames:2011xi}. We include as the original result for this section the case where $ \psi_n <0 $.
In Section 3, new features were introduced. Firstly, the stability of the black hole was analyzed considering thermodynamic aspects by the analysis of the regions of temperature positivity and heat capacity of the system, after this analysis, a comparison was made between Hawking and local formalisms.
We also include in this version the stability analysis by comparing the heat capacity and the temperature of the black hole. We also inspect, in this Section, the case with $\psi> 0$, where use is made of the local prescription, also valid for the case in which $\psi <0$.  In Section 4, we consider the thermodynamical properties of the black hole in an $f(R)$ global monopole space-time. Finally, in the Section 5, we cast our Closing  Remarks, discuss the stability of our solution and analyze the particular case of $n=2$ in more details; in this special case, the parameter $\psi_2$ can be related with a positive cosmological constant, that, in the presence of the monopole presence exhibits a deficit solid angle.

\section{Field Equations Solution  for the $f(R)$ Gravity in the Metric Formalism: A Review}

In this Section, for the sake of understanding, we make a small review of the solution of the field equations for an $f$(R)-theory with a spherically symmetric space-time obtained in \cite{Carames:2011xi}. In this reference it is  shown that it is possible to find a black hole solution in a global monopole region by using an $f$(R) modification of the GR gravity within the weak field approximation.  For an $f$(R) theory, the action associated to the matter field coupled with gravity is given by:
\begin{equation}
S = \frac{1}{2\kappa}\int{d^4x\sqrt{-g}f(R)} + S_m(g_{\mu\nu}, \psi),\label{action1}
\end{equation}
where $\kappa = 8 \pi G $, $G$ is the Newton constant, $ g$ is the determinant of the metric $g_{\mu \nu}$ with $\mu , \nu = 0, 1, 2, 3 $, $R$ is the curvature scalar,  $S_m $ is the action associated with the matter fields and $f(R)$ is an analytic function of the Ricci scalar. In this model, the Ricci scalar in Einstein -Hilbert action is replaced by $f(R)$. The monopole which introduces  an angular deficit in the space-time metric, gives us some interesting effects that we shall discuss in the sequel. Here, we assume that the Christoffel symbol is a function of the metric, its derivatives and its inverse. The general form of the time-independent metric with spherical symmetry in (1+3) dimensions  is given by
\begin{equation}
ds^2= B(r) dt^2 - A(r)dr^2 - r^2 (d\theta^2 + \sin^2\theta d \phi^2)   \label{metric1}
\end{equation}
The $g_{\mu \nu}$ field equations read as below::
\begin{equation}
R_{\mu \nu} F(R) - {1\over 2}f(R)g_{\mu \nu} - ( \nabla_{\mu}\nabla_{\nu}-g_{\mu \nu}\Box)F(R) = \kappa T_{\mu \nu}\label{freq}
\end{equation}
with $ F(R) = {df(R)\over dR} $, and $ \Box$ is the usual notation for covariant D' Alembert operator $ \Box \equiv \nabla_{\mu} \nabla^{\mu}$.
The only term associated to the matter action is related to the global monopole described by the Lagrangian density
\begin{equation}
{\cal L} = \frac{1}{2}(\partial_{\mu}\phi^a)(\partial^{\mu}\phi^a ) - \frac{1}{4}\lambda(\phi^a \phi^a - \eta^2)^2,
\end{equation}
where $\lambda$ and $\eta $ are the monopole field parameters and the triplet field that will result in a monopole configuration that can be described by $ \phi^a = \eta \frac{h(r)}{r^2}x^a $, with $ a = 1, 2, 3 $ and $x^a x^a = r^2 $.
     The function $h$(r) is dimentionless and is constrained by the conditions $h(0) = 0$ and $h(r>\eta) \approx 1$ \cite{Barriola:1989hx}. The energy-momentum tensor associated with that field configuration is
\begin{eqnarray}
T_{t}^{t} &=&  T^r_r \approx \frac{\eta^2}{r^2} \nonumber\\
T^\theta _\theta & = & T^\phi_\phi =0
 \label{energmoment}
\end{eqnarray}
We can now focus on the gravitational effects of the global monopole in the $f(R)$ approach to gravity, by taking the trace of the field equations (\ref{freq}) in the presence of the monopole\cite{Carames:2011uu,Multamaki:2006zb}

\begin{equation}
f(R) = {1 \over 2} R F(R) + {3\over 2} \Box F(R) - {\kappa \over 2} T\label{frcontract}
\end{equation}
where $F(R) = {df(R) \over dR}$. By replacing (\ref{frcontract}) in (\ref{freq}), we have
\begin{equation}
F(R) R_{\mu \nu} - \nabla_{\mu} \nabla_{\nu} F(R) - T_{\mu \nu} = g_{\mu \nu} \, C \label{solutionequation}
\end{equation}
where the quantity $C= {1\over 4}(F(R) \,R - \Box F(R) - \kappa T)$ is a scalar quantity.

In a spherically symmetric space-time, with the metric given by (\ref{metric1}) and with the energy-momentum tensor expressed by (\ref{energmoment}), we find, by virtue of (\ref{solutionequation}), that the equations may be written as
\begin{eqnarray}
2r F'' - \beta (r F' + 2 F) =0,  \label{frn}\\
4B(A\!-\!1)\!\! +\! r\Big( \! 2\! {F' \over F}\! -\! \beta \!\Big) \! (r B' \!-\! 2 B)\! +\! 2 r^2 B'' \! -\! {4 \alpha A B  \over F} =0\label{brn}
\end{eqnarray}
where $\alpha$ and $\beta$ are:
\begin{eqnarray}
\alpha &=& 2 \kappa \eta^2, \, \,  \mbox{with}  \, \, \kappa = 8\pi G \\
\beta &=& {(AB)' \over AB} \label{beta} \,\, .
\end{eqnarray}

 We make the assumption of time-independent  solutions, {\it i.e.}  $B = B(r)$ and  $A = A(r)$, which yields the metric as given in (\ref{metric1}). With these assumptions, we consider $F(R) = F(r)$ and $F'$ and $F''$ are the first and the second derivatives with respect to $r$, respectively.
The solution to these equations is the exact description of the global monopole in $f(R)$ theories.
This prescription has however an analytical solution only in the weak field approximation.  In this approximation, the field equations read as follows:
\begin{eqnarray}
F(R) = F(r) = 1 + \psi(r) \nonumber \\
B(r) = 1 + b(r),  \, \, \, A(r) = 1 + a(r)
\end{eqnarray}
with $\vert b(r)\vert $, $\vert a(r)\vert $ and $\vert \psi(r)\vert$ much smaller than one.  These redefinitions are used in equations (\ref{frn}) and (\ref{brn}), and with the help of
\begin{eqnarray}
{F' \over F} \sim \psi',& \, \, \, &  {F'' \over F} \sim \psi''\\
{A' \over A} \sim a', &\, \, \, & {B' \over B} \sim b' \label{ab}
\end{eqnarray}

the field equations (\ref{frn}) and (\ref{brn}) can then be written in a linear form as:
\begin{equation}
\beta \sim r \psi '' (r)\label{beta2},
\end{equation}
\begin{equation}
2a(r) - 2r\psi '(r) + r \beta + r^2 b '' (r) - 2 \kappa \eta^2 =0 ,\label{abequation}
\end{equation}
where  $\beta $ has been defined in (\ref{beta}). This is the solution to our equations in the weak field approximation for a metric with spherical symmetry in an $f$(R)-theory with $\frac{df(R)}{dR} = 1 + \psi(r)$ where R is the function of the radial coordinate, r.
Following in \cite{Carames:2011xi},  we assume that  the function
$\psi(r)$  is a power law-like function of the radial coordinate, namely $\psi(r) = \psi_{(n)} r^n$, where $\psi_n$ is a constant parameter in $r$.  Notice, however, that in our work it shows a dependence on $n$.
With this {\it ansatz}, we can find solutions for the equations above. Noting that equation (\ref{beta}) with (\ref{ab}) can be rewritten as:
\begin{equation}
a' (r) + b' (r) = n(n-1)\psi_n r^{(n-1)}
\end{equation}
we can solve (\ref{abequation}) for $b(r)$, obtaining:
\begin{equation}
b(r) = \frac{c_1}{r} - \psi_n r^n - \kappa \eta^2. \label{linb}
\end{equation}

We define the integration constants  $c_1 = -2GM$ and, for convenience, we take $c_2 =0$. This term corresponds to the cosmological constant term. Equations (\ref{beta}) and (\ref{beta2}) yield the relation:
\begin{equation}
A(r)B(r) = a_0 e^{(n-1)\psi_n r^n}.\label{AB}
\end{equation}
Defining the integration constant $a_0 = 1$, we then have the full form of the metric as
\begin{equation}
ds^2 = B(r) dt^2 - e^{(n-1)\psi_n r^n}B(r) ^{-1}dr^2 - r^2d\Omega^2, \label{metricfinal}
\end{equation}

\begin{equation}
B(r) = 1- \kappa \eta^2 - \frac{2GM}{r} - \psi_n r^n  \label{B}
\end{equation}

\begin{figure}[htb]
\includegraphics[width=8cm, height=6cm]{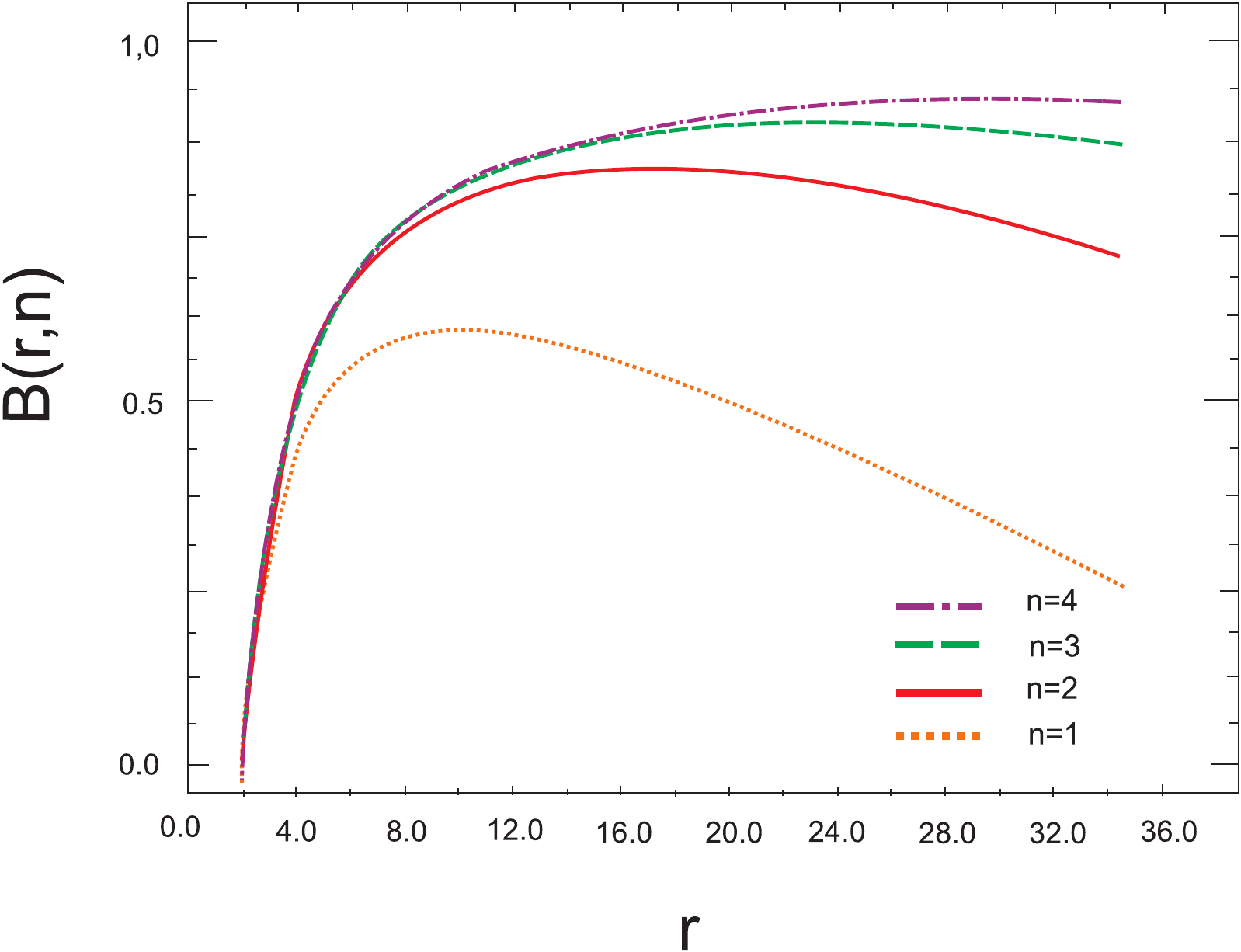}
\includegraphics[width=8cm, height=6cm]{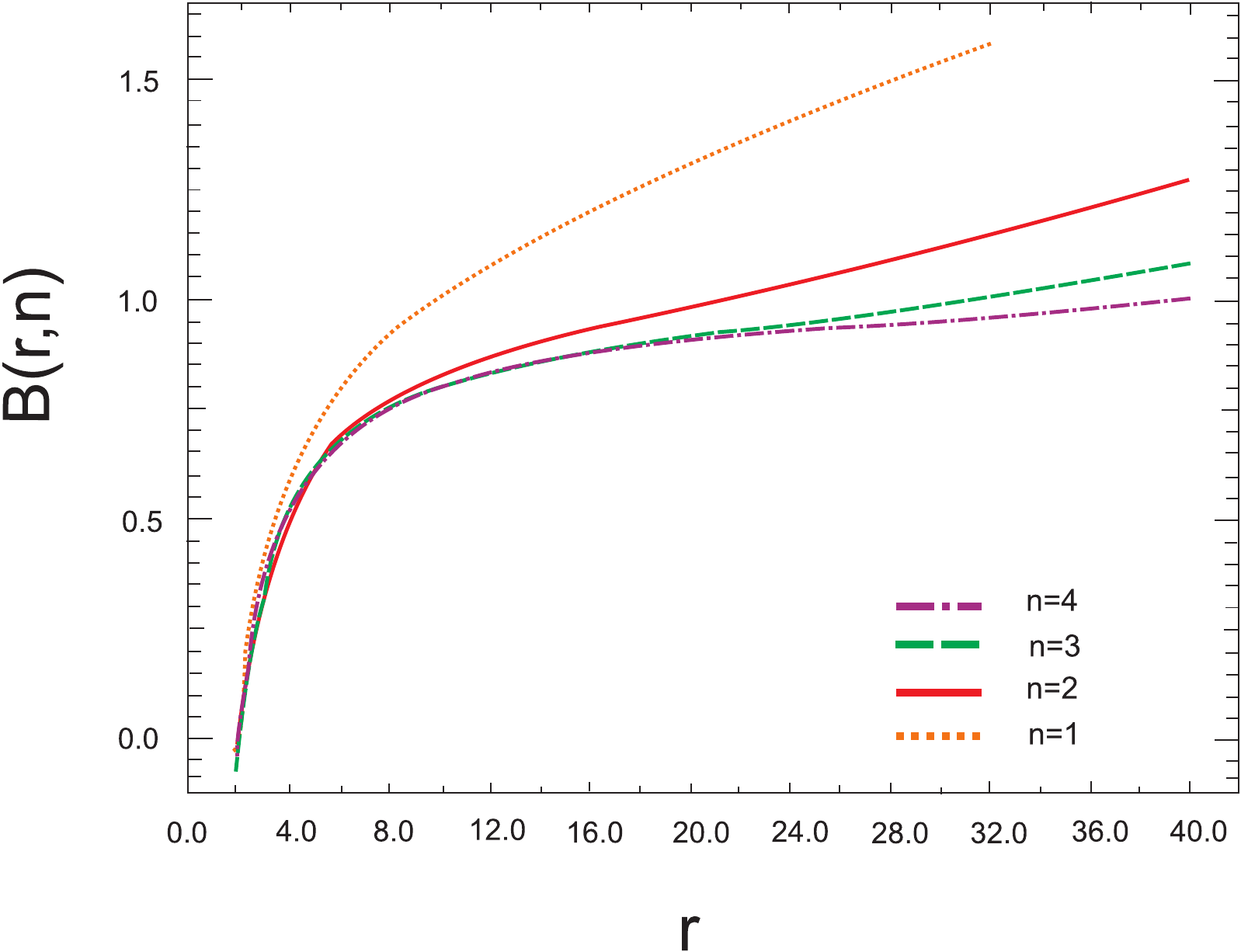}
\caption{Graphs for B(r, n) in function of position for  each n- orders going from 1 to 4, where $GM=1$ and $\kappa \eta^2 = 0.3 \cdot 10^{-5}$. In the figure a) $\psi_n =0. 2 \cdot 10^{-2n} $ and  b)    $\psi_n =- 0. 2 \cdot 10^{-2n} $   }
 \label{Figure1}
 \end{figure}

In Figure \ref{Figure1},  we plot the function $B(r,n)$
for $n$ varying from 1 to 4. It can be seen, in the case of the figure on the left, with $ \psi_n> 0 $, that the function $ B (r, n) $ grows to a value and then decreases for all orders of n. In the case $ \psi_n <0 $, the function transits to ever increasing values.

Considering the mass term as cosmologically relevant is equivalent to adopting the scenario with a black hole in the space-time of a global monopole  \cite{Barriola:1989hx}. To confirm that our solution is actually a black-hole-type solution for all degrees $n$ of the $\psi(r) $-function, we have to find if all of them have an event horizon.
     To do this, we have to keep in mind the approximations that we have used to solve the field equations, namely $h \approx 1 $, which we have employed to define the energy-momentum tensor outside the core of the global monopole, and $|\psi(r) |<< 1 $. These approximations require that we restrict our analysis to a region defined by:
\begin{equation}
\delta < r < \frac{1}{\vert \psi_n r^n \vert} \label{rlimits}
\end{equation}
where $\delta \approx (\lambda\eta^{1/2})$ is the order of magnitude of the monopole's core.

In the following Sections, we shall concentrate on the cases where $\psi_n <0$ and $\psi_n> 0$ in the limit given by (\ref{rlimits}). These solutions do not contemplate the asymptotic case when $n> 0$; however, its analysis is important if we wish to consider regions closer to the black hole.

\section{The Thermodynamics of the Black Hole in an $f(R)$ Global Monopole for Generic $n$.}

In this Section, the thermodynamic behavior is reported for two cases. In one of the cases, with $\psi_ <0$, the stability of the BH is ensured for all powers; the other case, where $\psi> 0$,  the technique of local thermodynamics has been adopted to analyze the stability of the BH .

\subsection{The Hawking Thermodynamics and the Power Law Series for a Generic n }

In this Sub-section, we shall be considering the thermodynamical behavior of a black hole in a region with a topological defect, by following the Hawking procedure.
We can analytically  with prove that the metric has an event horizon for the solution of an $(n + 1)$-degree equation:
\begin{equation}
1- \kappa \eta^2 - \frac{2GM}{r_{H}} - \psi_n r^n_{H}  = 0.\label{eqrh-mass}
\end{equation}

We have written the parameter $\psi_n $  as a power law-like function of radial coordinate for a generic n.  This is essential to respect the condition (\ref{rlimits}) without narrowing  the region that we are going to analyze. Considering only the region where $|\psi (r)| << 1 $ implies that the other roots of equation (\ref{eqrh-mass}) are out of the region where our solution is valid. In addition to the limit for $ \psi_n $ and $ \kappa \eta ^ 2 $, used in the previous graphs, the black hole (BH) stability was analyzed in detail considering the functions of temperature, heat capacity and its phase.

From  (\ref{eqrh-mass}),  we get  the dependence of the energy (GM) on the horizon $r_H$
 \begin{equation}
 GM = {1 \over 2} \Big[ (1-\kappa \eta^2)r_H - \psi_n r^{n+1}_H  \Big ]\label{mass2}
 \end{equation}

The Hawking  temperature can be readily obtained from the GM expression, which corresponds to the energy of the black hole. The Hawking temperature is, therefore, the  derivative of the energy with respect to the entropy of the black hole \cite{Faraoni:2010yi}

\begin{equation}
\label{entropia}
S =F(R) A/4\equiv {1 \over 4 }( 1+ \psi_n r_H^n)\pi r_H^2 \, ,
\end{equation}
where $A$ is the area of the event horizon.

\begin{figure}[htb]
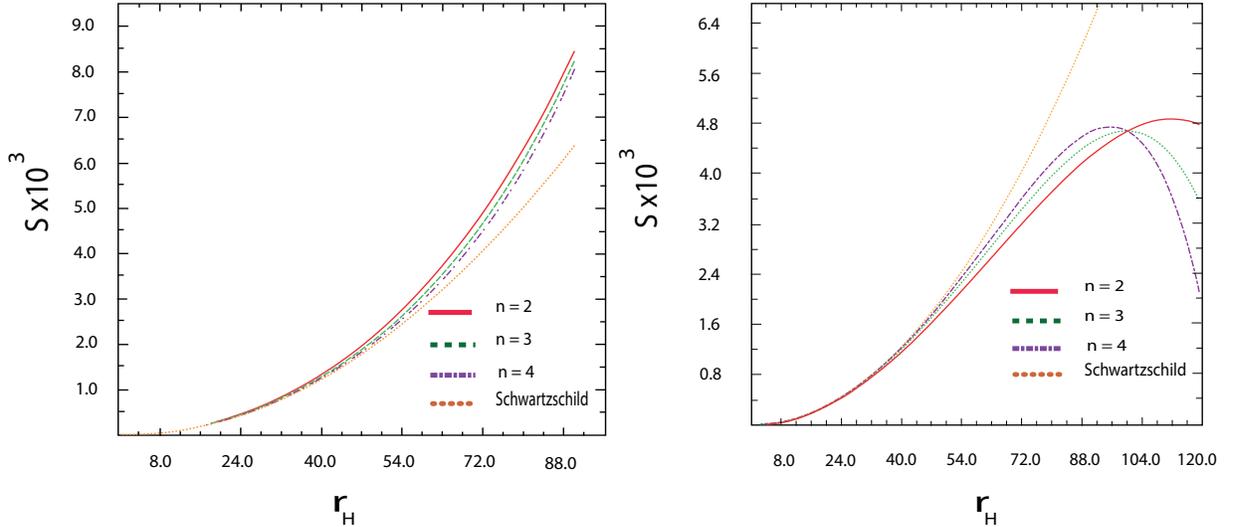

\includegraphics[width=8cm, height=7cm]{figSa.pdf}
\includegraphics[width=8cm, height=7cm]{figSb.pdf}
\caption{Graph for $S(r_H)$ for n going from 2 to 4 and Schwarzschild black hole with global monopole ,  $\kappa \eta^2 = 0.3 \cdot 10^{-5}$ where in left  panel  $\psi_n = 0. 4 \cdot 10^{-2n} $ and in the right  panel   $\psi_n = -0. 4 \cdot 10^{-2n} $}
\label{Figure2}
 \end{figure}

In Figure \ref{Figure2}, which shows graphically the behavior of the entropy with the radius of the horizon, it indicates that  there are critical points for different values of n given by the expression to $r_H$:
\begin{equation}
r_{H_{max}} = 2^{1/n} \left(    -\psi_n (n+2)  \right)^{-1/n} \label{rmax}
\end{equation}
In this expression it can be easily verified that $r_{H_{max}}$ applies only when $\psi <0$ ,  as can be seen by the graphs of Figure \ref{Figure2}. This fact has given us an important result which was related to the critical range between allowed and forbidden region, but  throughout this work will be analyzed if this transition is within the limit of validity of the solution found.

\begin{figure}[htb]
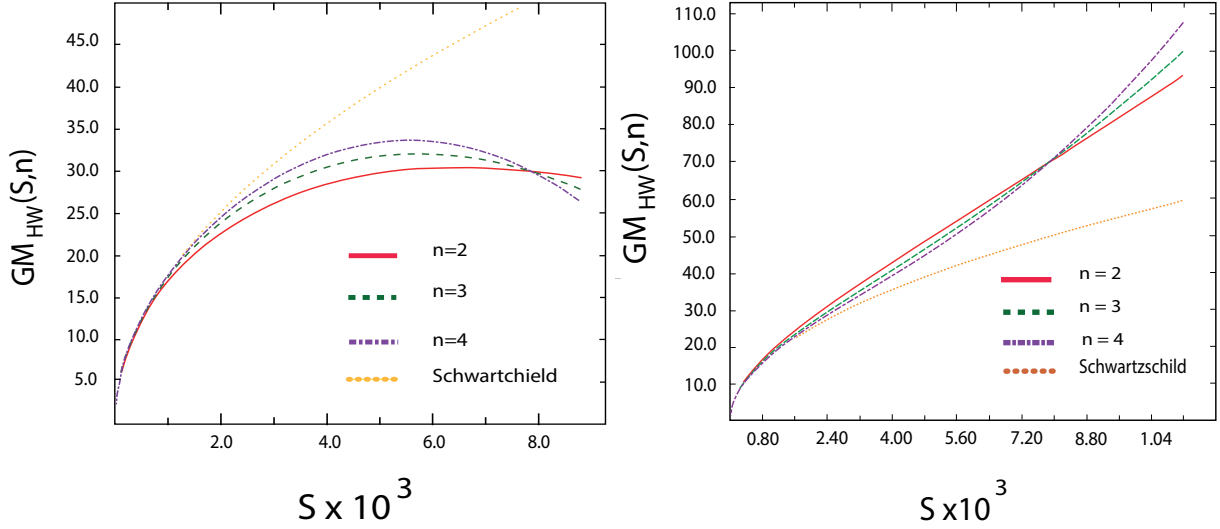

\includegraphics[width=8cm, height=7cm]{fig2-1a.pdf}
\includegraphics[width=8cm, height=7cm]{fig2b.pdf}
\caption{Graphs for $GM_{HW}(S, n)$ for n going from 2 to 4  and the Schwarzschild black hole case,  $\kappa \eta^2 = 0.3 \cdot 10^{-5}$ and $\psi_n = \pm 0. 4 \cdot 10^{-2n} $. }
\label{Figure3}
 \end{figure}

In  the Figure \ref{Figure3},  we plot the energy in terms of the entropy of the black hole. It is interesting to investigate how the analogous relations for gravity and thermodynamics become; with the development of a quasi-classical theory \cite{Bardeem73, Bekenstein:1973ur,Hawking74}, in a widely accepted relation in which the area of the  black hole corresponds in fact to its entropy and consequently, its superficial gravity is responsible for measuring the temperature that the radiation emitted by the black hole will have. Although we are far from testing these results experimentally, the apparent lack of inconsistencies in the theoretical model and its application to several cases of interest [4] make it widely accepted.

The prescription for the Hawking's temperature is given by the second low $T= {d GM \over dS} $ given by

\begin{equation}
T_{HW} = {dGM \over dr_H} \Big( {dS\over dr_H} \Big)^{-1}
\end{equation}
where

\begin{eqnarray}
{dS \over dr_H} & = &  {\pi r_H \over 2}\Big[1+ (1+ {n \over 2})\psi_n r_H^n  \Big] \\
{dGM_{HW} \over dr_H} &= &{1\over 2} \Big( 1- \kappa \eta^2 - (n+1) \psi_n r_H^n\Big)
\end{eqnarray}

  \begin{eqnarray}
T_{HW}(\eta, \psi_n) =  \frac{1}{\pi \,  r_H}{\Big[1 - \kappa \eta^2 - (n+1)\psi_n r_H^n\Big] \over \Big[ 1+(1+{n \over 2}) \psi_n r_H^n\Big]}, \label{TH}
\end{eqnarray}
which, for $n = 1$, recovers the result obtained in \cite{Man:2013sf}, and, for $\eta = 0$, $\psi(r) = 0$, the Schwarzschild result is reproduced. In this work, we concentrate in the case $n>1$  with a Schwarzschild-like 
solution in presence of a global monopole.

\begin{figure}[htb]
\includegraphics[width=8cm, height=7cm]{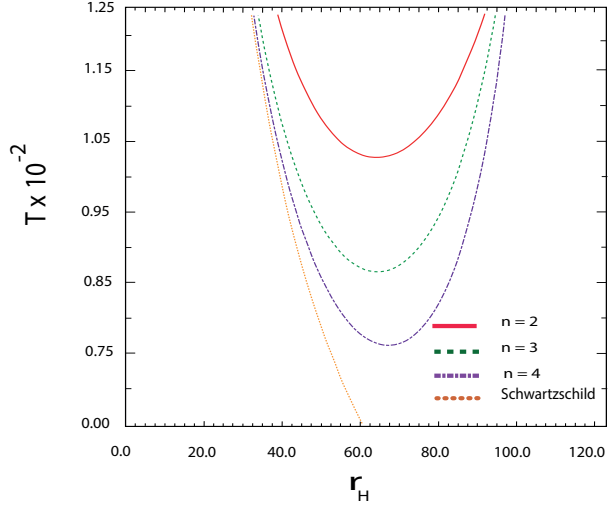}
\caption{Graph for the Hawkings  temperature, $ T$ in function of the horizn radio $r_H $, with  $\kappa \eta^2 = 0.3 \cdot 10^{-5} $ and n going from 2 to 4, with Schwarzschild black hole with global monopole $\psi_n =-0. 4 \cdot 10^{-2n}  $. }
\label{Figure4}
 \end{figure}

Figure \ref{Figure4} shows that, for all values of $n$, there are two types of the black hole temperature: one with the temperature decreasing with the horizon, $r_H$, and the other with the temperature growing with the horizon.
The minimum point between these two regimes is calculated in this model by the derivative of the Hawking temperature as a function of the horizon radius ${d T_{HW}\over r_H}=0$. The expression resulting from this calculation is a simple polynomial of degree n, for $ n \neq1 $. In this graph it can also be seen that the temperature behavior in this region has the same shape and transition. This fact has not been applied to the whole region of space.

Despite the positivity of the temperature, to analyze the stability of the black hole, it is necessary to verify the prescription for the heat capacity. Using the laws of thermodynamics for black holes, it has been found that it is possible to obtain the heat capacity of the black hole by considering the entropy of the system, {\it i.e.}, $ dE = T dS $. In the case where the volume is constant, the energy, E, can be related to the mass of the black hole.

We shall now look for the heat capacity, which can be calculated from the energy. The expression is obtained as follows:
\begin{equation}
C_{HW} = \Big(\frac{d GM}{d T_{H}}\Big)_r =\Big[\frac{d GM_{H}}{d r_H} \cdot  \Big(\frac{d T_{H}}{d r_H}\Big)^{-1}\Big]_r
\end{equation}

\begin{eqnarray}
u&= {dGM_H \over dr_H}&= {1\over 2} \Big(1-\kappa \eta^2 -(n+1) \psi r_H^n \Big)\nonumber \\
v&= { d T_H \over d r_H}&= -{1 \over 4 \pi [ 1+(1+{n \over 2})\psi_n r_H^n]^2 r_H^2}  \Big\{1-\kappa \eta^2 +(n^2-1)\Big[ 1+  {(1+{n \over 2}) \over (n-1)}(1 - \kappa \eta^2 - \psi_n r_H^{n}) \Big] \psi_n r_H^{n} \Big\}\nonumber
\end{eqnarray}

\begin{equation}
C_H= {u\over v}
\end{equation}

By considering equations (\ref{TH}) and (\ref{mass2}), we obtain the expression:
\begin{equation}
C_{HW} = -2 \pi r_H^2{  \Big[1-\kappa \eta^2  -(n+1)\psi_n r_H^n ] \Big[1+(1+{n \over 2} ) \psi_n r_H^2\Big]^2     \over   1-\kappa \eta^2 +(n^2-1)\Big[ 1+  {(1+{n \over 2}) \over (n-1)}(1 - \kappa \eta^2 - \psi_n r_H^{n}) \Big] \psi_n r_H^{n} }
\end{equation}

\begin{figure}[htb]
\includegraphics[width=8cm, height=7cm]{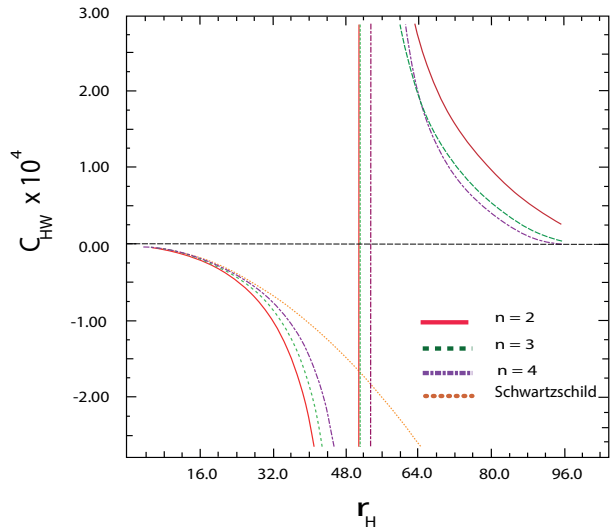}
\caption{Graph for heat capacity, $ C$, with  $\kappa \eta^2 = 0.3 \cdot 10^{-5} $ with  $\psi_n= - 0.4 \cdot 10^{-2n} $  and panel right. }
\label{Figure5}
 \end{figure}

In Figure \ref{Figure5}, we see the heat capacity  plotted for the classical case and the $n = 2$ to  $4$  and compared  with the Schwarzschild  black hole.

By inspecting Figures \ref{Figure4} and \ref{Figure5}, one can discuss the the validity regions of our solution of the BH in terms of the temperature and heat capacity. It can be seen that, for the region $\delta \leq r_H < r_H^* $, the temperature is positive but the specific heat is negative. This region is therefore prohibited because it has a negative heat capacity because it is not in accordance with the laws of thermodynamics.
There is a discontinuity in the heat capacity equation at the point $r_H = r ^ * _H$, which can be seen in the graph of Figure 5. This discontinuity is the point where the specific heat changes behavior, becoming positive.  In this region both temperature and heat capacity are positive, with the possibility of obtaining an allowable solution for the black hole.


Due to the maximum point given by the equation (\ref{rmax}) and which can be seen graphically in Figure 2, we have another phase transition in this system. For $r_H> r_{max} $ the black hole becomes unstable again, although the heat capacity is positive. The heat capacity at this point goes from decreasing to the radius of the horizon, to increasing with the radius of the horizon, and this destabilizes the BH presenting negative temperatures, in this way this region becomes forbidden as we can saw in Figure 6. In the next section we will discuss the local case, where we will compare the local temperature with the Hawking temperature, this comparison gives us new results.

\begin{figure}[htb]
\includegraphics[width=8cm, height=7cm]{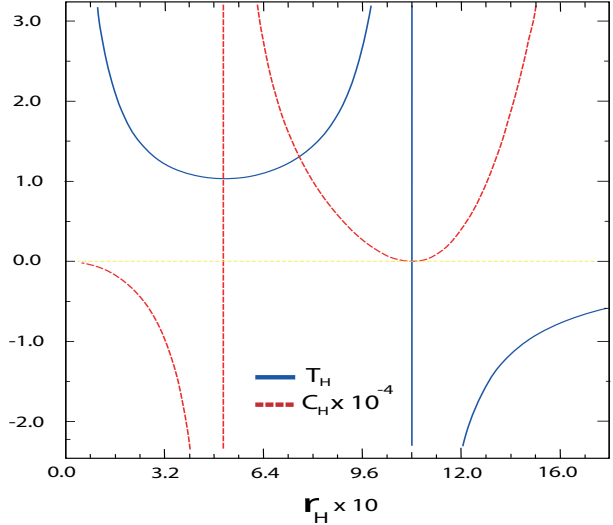}
\caption{Graph for  temperature and heat specific, $ C$ in function of the  horizon ratio for n=2, with  $\kappa \eta^2 = 0.3 \cdot 10^{-5} $ and   $\psi_n= - 0.4 \cdot 10^{-2n} $ . }
\label{Figure6}
 \end{figure}

In the graph of Figure 6 it is easy to see that the transition is located exactly at the maximum point of the entropy curve given by the expression (\ref{rmax}).

It may be verified that, whenever $\psi_n> 0$,  both regions are prohibited, because, for a positive temperature, the heat capacity is negative, and when the heat capacity is positive, the temperature is negative. Then, in the case of $ \psi_n> 0 $  it is necessary to work with  another prescription that, in this paper, we take as the local prescription.

\subsection{The Local Thermodynamics of the Black Hole for a Generic $n$.}

In the previous Section, we have treated the thermodynamical properties by considering the Hawking formalism, where it is possible to investigate the properties in the case  $ \psi_n <0 $. In the present Section, we go through a formulation where it is possible to discuss the stability conditions independently of the signal  $ \psi_n$.
We have nevertheless to recall the limitations imposed in the metric analysis (\ref{metricfinal}). The solution that led us to this configuration was found by using weak field approximations, which consists in disregarding any cross-terms. In addition, the metric (\ref{metricfinal}) is only valid locally, for a region $ \delta <r <1 / | \psi_n | $, also we analyzed in Hawking's case, which forces us to work with local thermodynamical quantities. The Schwarzschild metric does not present such limitations, but for the purposes of comparison, in this Section the local thermodynamic quantities are going to be derived. We use the prescription given in \cite{Tolman:1930zza} to obtain the local temperature of the black body radiation:

\begin{equation}
T_{loc} = {T \over \sqrt{g_{00}}}
\end{equation}
with this relation the temperature was obtained as a function of the position as

\begin{eqnarray}
T_{loc} = \frac{1}{\pi}{\Big[\frac{1 - \kappa \eta^2}{r_H} - (n+1)\psi_n r_H^{n-1}\Big] \over \Big[ 1+(1+{n \over 2}) \psi_n r_H^n\Big]} \sqrt{ \!\frac{r}{\psi_n r_H^{n+1}\!\! \!-\! r_H(1 \!- \! 8\pi G \eta^2) \! \!+ \!r(1\! - \! 8\pi G \eta^2) \!- \!\psi_n r^{n+1}}}\, .\label{Tloc}
\end{eqnarray}
     The local temperature for $n=2$ to $4$ and  the Schwarzschild black hole is plotted in Figure 7 as a function of the event horizon where we fixed the position at $r = 10$. The graph is zoomed to show the slight variation difference between the minimum values of the temperature for different values of the degree $n$.

\begin{figure}[h]
    \centering
    \includegraphics[width=8cm, height=7cm]{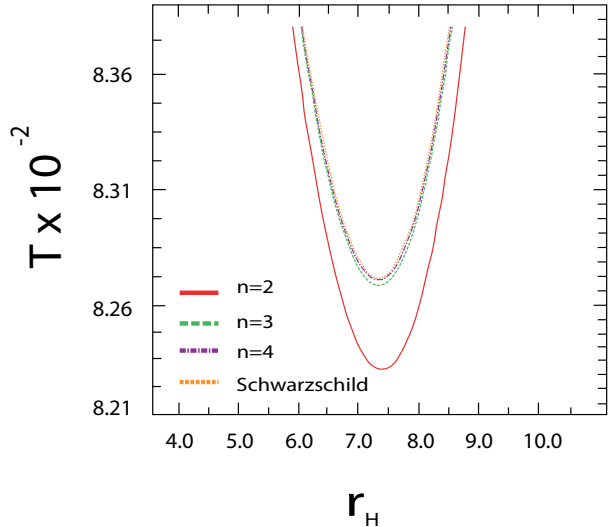}
    \caption{Graph for $T_{loc}(r_H)$ for $n= 2,  3, 4$ and Schwarzschild, for $r=10$.}
    \label{Figure7}
  \end{figure}

An important characteristic to observe at this temperature is the existence of a minimum in its variation with respect to the event horizon. This minimum can be called the critical temperature [29], which would be the minimum temperature in the region around the black hole. In Figure 7 the local temperature curves for n = 2,  3, 4 and the Schwarzschild  were calculated numerically. Displays that, for all values of $n$, there are two possible black holes: one for which the temperature diminishes with the horizon,  to $r_H > r^*_H$, and the other with the temperature growing with the horizon, to $r_H> r_{min}$.  These phase transitions were analyzed in more detail in the local case in connection with the heat capacity formalism.

It is important to notice, from this and the following graphs, that we fix a position for the measurement of the temperature and we are analyzing how the temperature in that position varies as the event horizon increases. That means, we are finding what size the black hole has at the moment in which the temperature in the position r = 10 is minimal. In this sense, we see that as the horizon approaches the chosen position the temperature decreases rapidly to the minimum and then increases asymptotically when it approaches r = 10.

From the first law of thermodynamics $ dGM_ {loc} = T_ {loc} dS $, the thermodynamical local energy, $ GM_{loc} $, can be derived. Following the prescription discussed in last section, we can assume that  $S =F(R) A/4\equiv {1 \over 4 }( 1+ \psi_n r_H^n)\pi r_H^2$, where A is the area of the event horizon.

\begin{equation}
GM_{loc} =     E_0 + \int_{S_0}^{S} T_{loc} dS = E_0 + {\pi \over 2}\int_{r_H (M=0)} ^{r_H(M)}T_{loc}(\xi) \Big[1+ (1+{n\over 2}) \psi_n\, \xi^n  \Big]\, \xi  \,d \xi
\end{equation}
where we choose $ E_0 = 0 $ for simplicity. Using the expression for the local temperature in the integral using
\begin{equation}
r_H(M=0) = [(1-\kappa \eta^2)/\psi_n)]^{-1/n}
\end{equation}
With  the integration constants conveniently, we get
\begin{eqnarray}
GM_{loc} = r\sqrt{(1 - 8\pi G \eta^2 - \psi_n r^n)} -\sqrt{r}\sqrt{\! r(1 \!- \!8\pi G \eta^2 \!-\! \psi_n r^n) \!-\! r_H(1 \!- \!8\pi G \eta^2 \!- \!\psi_n r_H^n)}\ ,\label{Eloc}
\end{eqnarray}
 In Figure 8, we plot the Energy as a function of the entropy.  It can be seen that, in the local framework, the energy and the temperature $T$ are both positive.

\begin{figure}[htb]
\includegraphics[width=9cm, height=7cm]{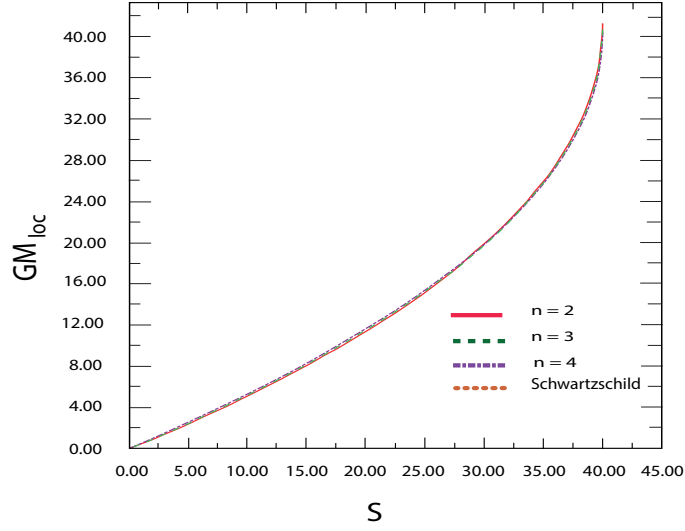}
\caption{Graph for $GM(S, n)_{loc}$ for n going from 2 to 4 and and Schwarzschild r= 15,  $\kappa \eta^2 = 10^{-5}$ and $\psi_n = 0. 4 \cdot 10^{-2n} $. }
\label{Figure8}
 \end{figure}

We are now going to look for the local heat capacity, which can be calculated from the energy. The expression is obtained as follows:
\begin{equation}
C_{loc} = \Big(\frac{d GM_{loc}}{d T_{loc}}\Big)_r =\Big[\frac{d GM_{loc}}{d r_H} \cdot  \Big(\frac{d T_{loc}}{d r_H}\Big)^{-1}\Big]_r \label{Cloc}
\end{equation}
The subscript is there to show that we are calculating this quantity in a fixed position. We work with Eq. (\ref{Cloc}) and give us
\begin{eqnarray}
{dGM_{loc} \over dr_H} & = &{\sqrt{r} \over 2} \Big(1 - 8 \pi G \eta^2 - (n+1)\psi_n r_H^n \Big) \nonumber  \\
 & & \times \Big[r(1 - 8 \pi G \eta^2 - \psi_n r^n) - r_H(1 - 8 \pi G \eta^2 - \psi_n r_H^n)\Big]^{-1/2}, \\ \nonumber \\
{ dT_{loc} \over dr_H } &=& v \sqrt{r} \times \Big[r(1 - 8 \pi G \eta^2 - \psi_n r^n) - r_H(1 - 8 \pi G \eta^2 - \psi_n r_H^n)\Big]^{-1/2}\nonumber \\
& & +{T_H \over 2} \sqrt{r} \Big( 1-8 \pi G \eta^2 - (n+1) \psi_n r_H^{n+2} \Big) \nonumber \\
& & \times \Big[r(1 - 8 \pi G \eta^2 - \psi_n r^n) - r_H(1 - 8 \pi G \eta^2 - \psi_n r_H^n)\Big]^{-3/2}
\end{eqnarray}

\begin{figure}[htb]
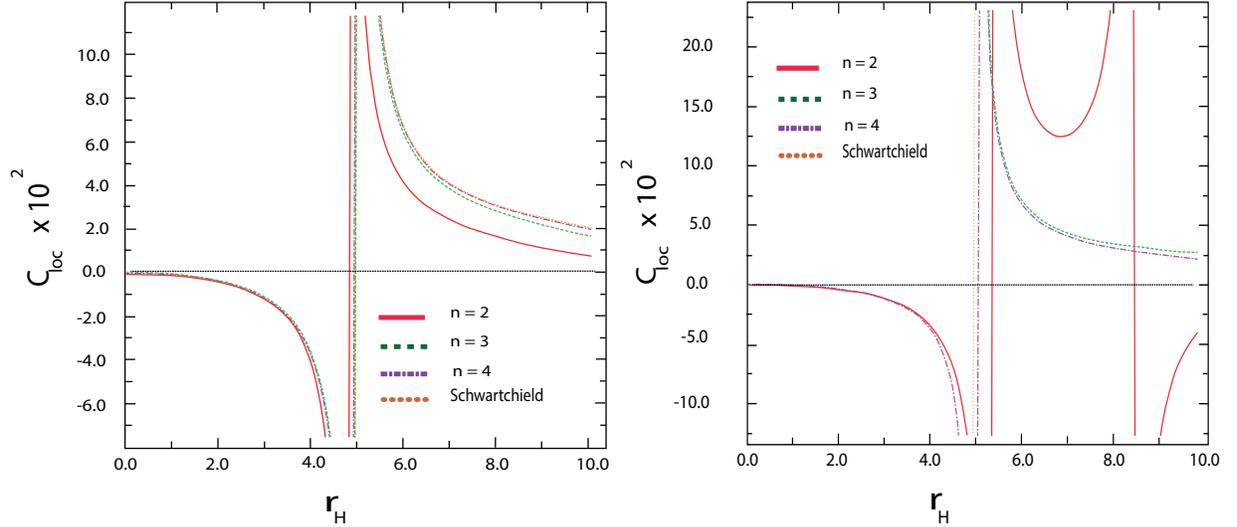

    \centering
     \includegraphics[width=8cm, height=7cm]{fig9a-1.pdf}
     \includegraphics[width=8cm, height=7cm]{fig9b-1.pdf}
    \caption{Graph for $C_{loc}(r_H,n)$ for  the  $n = 2$, $n = 3$ and $ n=4 $  and the Schwarzschild for black hole cases, for fixed $r=10$ , $\kappa \eta^2 = 10^{-5}$ and $\psi_n =\mp 0. 4 \cdot 10^{-2n} $.}
    \label{Figure9}
  \end{figure}

  In Figure 9, we see the local heat capacity plotted for the classical case and the $n = 2$ to  $4$ and  the local form for the Schwarzschild black hole. It can be observed that for the local case, it is possible to obtain an analysis for the thermodynamics of black holes for both $\psi <0$ and $\psi> 0$ to r=10. It can be seen that the curves for heat capacity are similar for the case $n> 2$. In fact in both cases the r limit of truncates the function.

The transition from negative to positive values happens as the horizon is closer to the chosen position, $r$ also vanishes present only the positive part.
We can notice this feature more clearly by analyzing the heat capacity as a function of the position, with a fixed horizon.
     The expected negative heat capacity will be observed as $r$ grows and the transition to positive values happens even for bigger black hole. This transition is yet to be explained and it occurs even for the classical Schwarzschild case in the local analysis, but not in the most usual global calculation where it remains negative for every value of $r_H$. It can be observed that for the local case, it is possible to obtain case for $\psi> 0$, that to $ n> 2$ has been similar to case $\psi <0$, in the case of small r.

         \begin{figure}[h]
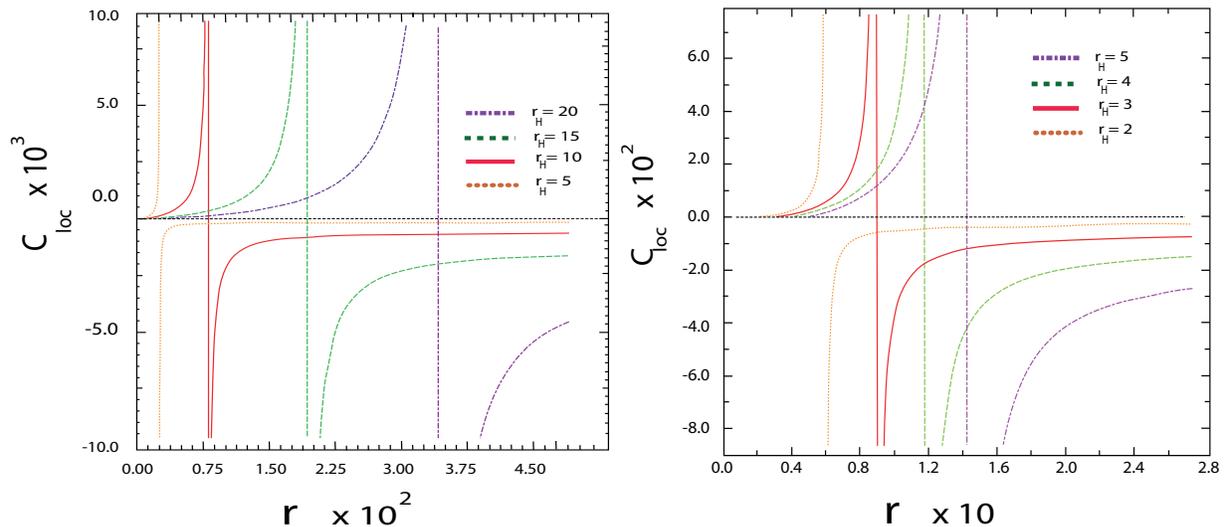

    \centering
    \includegraphics[width=8cm, height=7cm]{fig10a-1.pdf}
    \includegraphics[width=8cm, height=7cm]{fig10b-1.pdf}
    \caption{Local heat capacity for n= 2 as a function of the position, $\kappa \eta^2 = 10^{-5}$ and $\psi_n =- 0. 4 \cdot 10^{-2n} $.}
    \label{Figure10}
  \end{figure}

With Figure 10 it can be seen more easily that the function of the specific heat is stopped precisely at the point $r_{limit}$. Now we can analyze the case $n = 2$, where another phase transition appears where the specific heat changes direction and direction in case $\psi> 0$. It is found that when r is large this would happen for all n. However, when r is very large, the specific heat becomes totally negative and BH will not even form.

  Turning to the transition of the entropy phase, due to the correction introduced by the $ f (R)$  theory discussed in \cite{Faraoni:2010yi}, we can have two cases the first for $ \psi <0 $, which is analyzed in a shown in Figure 11  on the left to n=2. The Hawking temperature introduces a maximum limit for the horizon radius $ r_H $, which divides a permitted region with, $ T_{ HW}> 0$, of a forbidden region $ T_{ HW} <0 $. This maximum limit in the radius of the horizon $ r_H $ can be calculated using the Eq. (\ref{rmax}) and corresponds to the solid line of the graph.

       \begin{figure}[h]
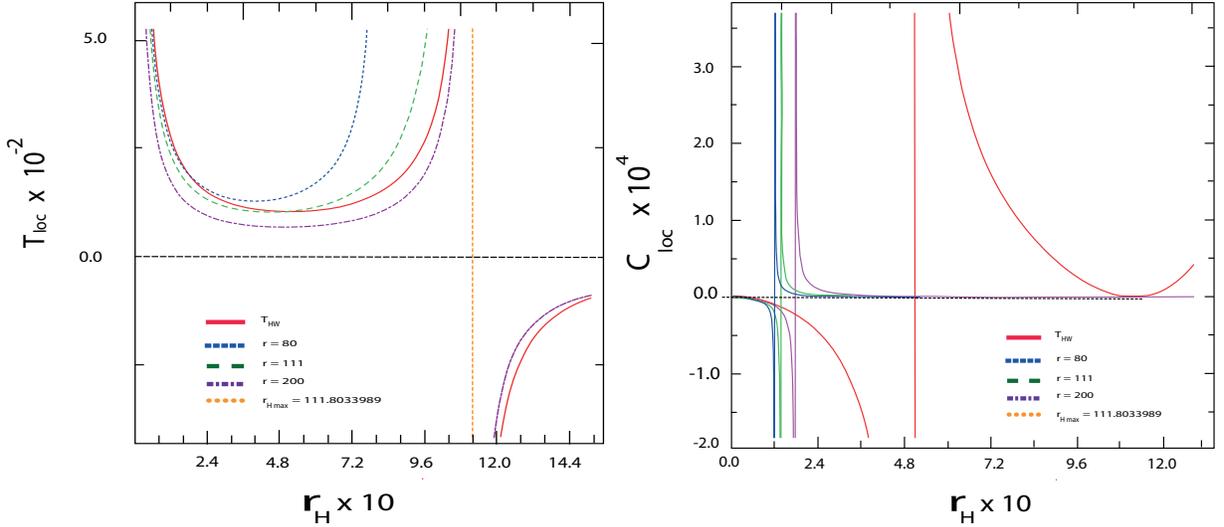

    \centering
    \includegraphics[width=8cm, height=7cm]{fig11a-1.pdf}
      \includegraphics[width=8cm, height=7cm]{fig11b-1.pdf}
    \caption{Local temperature  in the vicinity of $r_{H_{max}}$ for n= 2  in panel left with $\psi_n =- 0. 4 \cdot 10^{-2n} $ compared with Hawking temperature where $r_{H_{max}} = 111.8033989$ and in panel right  the heat capacity. }
    \label{Figure11}
  \end{figure}

  An interesting analysis is that the limit of $r_ {Max}$ that appears in the temperature of Hawking is also present in the local case. In the local case, this boundary also divides a permitted region of a forbidden region, and this happens when the position $ r = r_{H_{max}} $. When $r> r_{H_{Max}}$, the minimum temperature decreases, but the allowed region always has the same size as $ r_{H_{Max}}$.

This behavior can be seen in the graph of Figure 11 on the right, showing the value of $ r_{H_{Max}}$ for n = 2, considering different values of r. It can be shown that for $ r < r_{H_{ Max}} $  there is no forbidden region.

The existence of a $ r_H $ limit for black holes that has been studied in the case $ \psi <0 $ repeats to $ \psi> 0 $ existing a value of $ r_H$ that can not be passed $ r_{H_{Max}}$ . The Hawking temperature for $ \psi> 0 $ has no minimum point becoming negative, the $ r_{H_{max}}$ can be found when that temperature is zero by doing the zero expression (\ref{TH}) that gives us:
\begin{equation}
r_{H_{Max}} = \left( {1-\kappa\eta^2 \over \psi_n (n+1)} \right)^{1/2}
\end{equation}

In the local case we have a behavior similar to $\psi <0$ the maximum value for r is $r_{H_{Max}}$, however in this case there will always be a transition from the allowed region to the forbidden region, even when $r <r_{H_{Max}}$ in the vicinity of $r_{max}$. If we analyze the heat capacity we can observe that the validity limit to $r$ is very small  and far from $r_{r_{H_Max}}$ because the heat capacity became negative and the black hole is not stable.

\subsection{The Thermodynamical Phase Transitions for the Black Hole Systems}

We shall take care in this Section of the possible phase transitions of the system we are considering. The first one is the Hawking-Page phase transition.
It occurs whenever the function that gives the behavior of the Helmholtz free energy with the radius of the horizon has a root which is a minimum of this function.
\begin{eqnarray}
F_{HP}^{HW}|_{r_H = r_{HP}}& =& GM - T_{HP} S =  {1 \over 2} \Big[ (1-\kappa \eta^2) - \psi_n r^{n}_{HP}  \Big ]r_{HP} -
  {1 \over 4}( 1+ \psi_n r_{HP}^n)\pi T_{HP }  r_{HP}^2 =0 \nonumber \\
 \left( {\partial F_{HP}^{HW} \over \partial r_{H}} \right)_{r_H = r_{HP}}& = & {1 \over 2}\Big[ (1 -\kappa \eta^2 - (n+1) \psi_n r_{HP}^{n}\Big] - {\pi \over 2} T_{HP}\Big[ 1 + ({n\over 2}+1) \psi_n r_{HP}^{n} \Big] r_{HP}=0
  \end{eqnarray}
where $ T_{HP} $ is the Hawking-Page  temperature.
We find the radius and the Hawking-Page temperature as given below:

\begin{eqnarray}
T_{HP}^{HW}&=& \frac{1}{\pi \,  r_{HP}}{\Big[1 - \kappa \eta^2 - (n+1)\psi_n r_{HP}^n\Big] \over \Big[ 1+(1+{n \over 2}) \psi_n r_{HP}^n\Big]}\\
r_{HP} &= &\left[{n-{1 \over 2} \alpha (n+1) \pm {1 \over 2}\Delta^{1/2} \over \psi_n}   \right]^{1/n}  \\
\Delta &= &\alpha^2(n+1)^2 -4 \alpha (n^2+n+1)  +4(n^2+1)
\label{raiohawkingpage}
\end{eqnarray}

In Figure 12, the free energy question can be analyzed by considering the Hawking-Page phase transition showing its phase change for each value of n using Hawking's formalism. In the local case we can have this transition for both $\psi <0$ and $\psi> 0$. Repeating what happens in the case of the Hawking-Page radius, these values are very close, so we chose $n = 2$ to do the analysis as shown in Figure 13.

  \begin{figure}[htb]
\includegraphics[width=8cm, height=7cm]{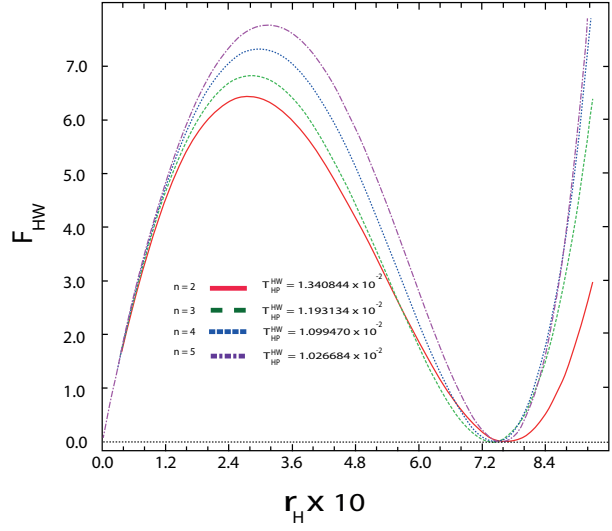}
\caption{Graph of $F_{HW}$ for the n-order going from 2 to 5, where $GM=1$, $\kappa \eta^2 = 0.3 \cdot 10^{-5}  $ and $\psi_n =-0.4 \cdot 10^{-2n}$ .}
 \label{Figure12}
 \end{figure}

  In Figure 13, we present the possible regimes for the free energy in the case $n = 2$. The same analysis applies for the others orders of n. The Hawking case that correspond to $ \psi_2 <0 $ was plotted in the panel of the left. In this case, we have a negative cosmological constant.
In the right plot, we have $ \psi_2> 0 $ that have similar results to $\psi_2<0$ in the local case, which can represents  when $\psi>0$ the case with a positive cosmological constant.

  \begin{figure}[htb]
\includegraphics[width=7cm, height=6cm]{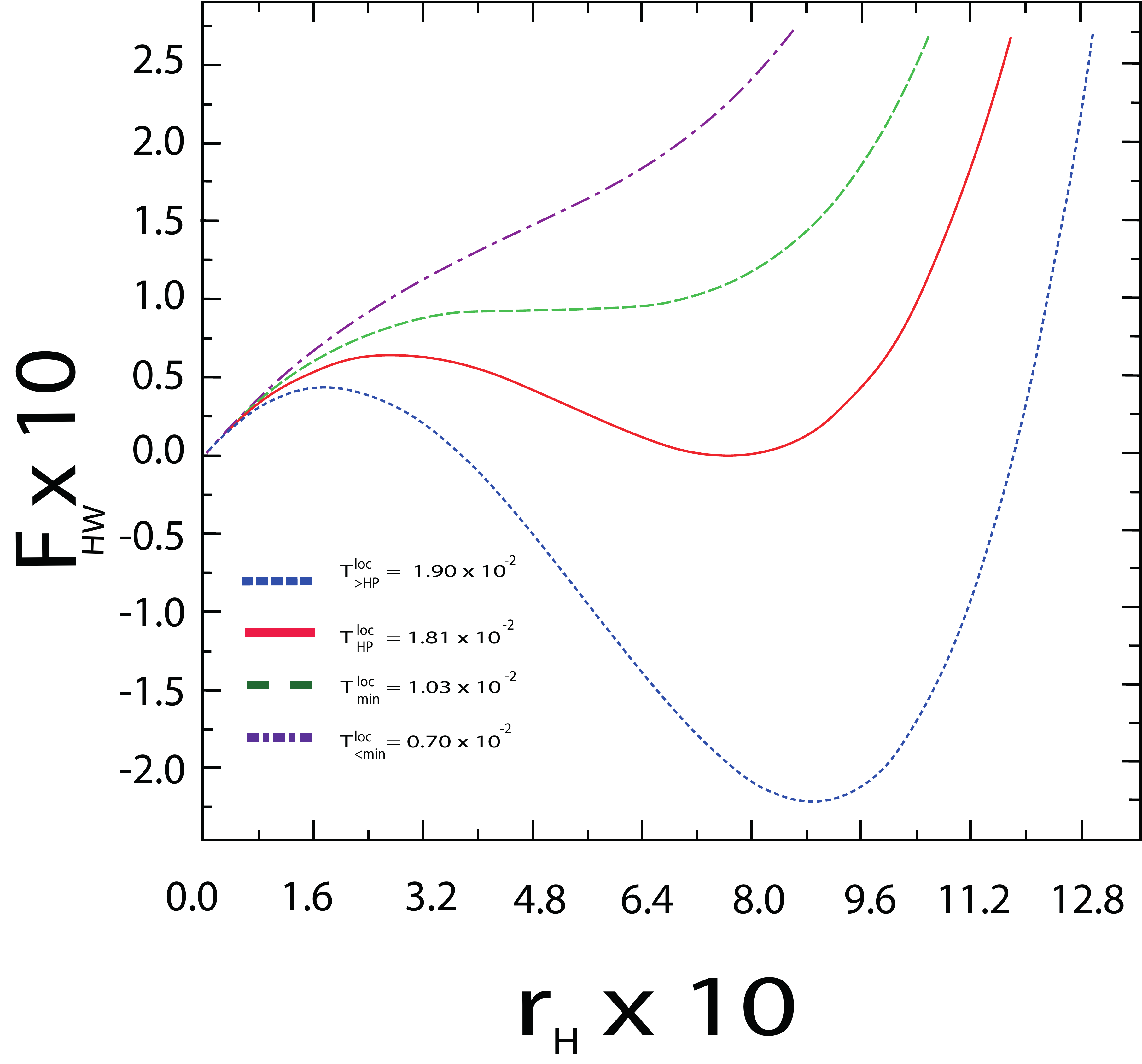}
\includegraphics[width=7cm, height=6cm]{fig13b-1.pdf}
\caption{Graph for F for  each n=2, $\kappa \eta^2 = 0.3  \cdot 10^{-5}$. The solid line is the Hawking-Page free energy, the dashed line is the minimal energy the dashed dotted line is the energy when $T>T_{HP}$ and the dotted line is the energy when $ T< T_{min} $ and   in a) the Hawking framework $\psi_n =-0.4 \cdot 10^{-2n}$   b) The local framework $\psi_n =0.4 \cdot 10^{-2n}$  .}
 \label{Figure13}
 \end{figure}

 The region of temperatures higher than the Hawking-page temperature of the Black Hole will always be stable. The minimum temperature puts into evidence a black hole's unstable region. In both cases, for $ T <T_{ min} $, where $ T_{ min}$ is represented by the dashed curve, we have a region of pure radiation. The solid curve represents the Hawking-Page temperature that gives us the Hawking-Page Phase transition. The  expression, for the minimum of the temperature in Hawking case, can be readily worked out and gives us

\begin{eqnarray}
T_{min}&=&{1\over 4 \pi }\Big( {1-\kappa \eta^2 \over r_{min}}-(n+1) \psi_2 r_{min}^{(n-1)} \Big)\\
r_{min}&=& \Big({-(1- \kappa \eta^2)\over (n^2-1) \psi_2} \Big)^{1/2}
\end{eqnarray}

As previously discussed, this work considers the boundary of the weak field approximation in a region in the vicinity of the BH with the defect. For this reason, we do not contemplate the asymptotic limits. However, whenever $n = 2$ and $ \psi $ is negative, the behavior of BH in the f(R)-model  considering the Hawking framework resembles the one which occurs in the case of the negative cosmological factor corresponding to the Anti-de-Sitter case. This case is extremely important for the study of the quantum behavior of gravitational systems.

  \begin{figure}[htb]
\includegraphics[width=9cm, height=7cm]{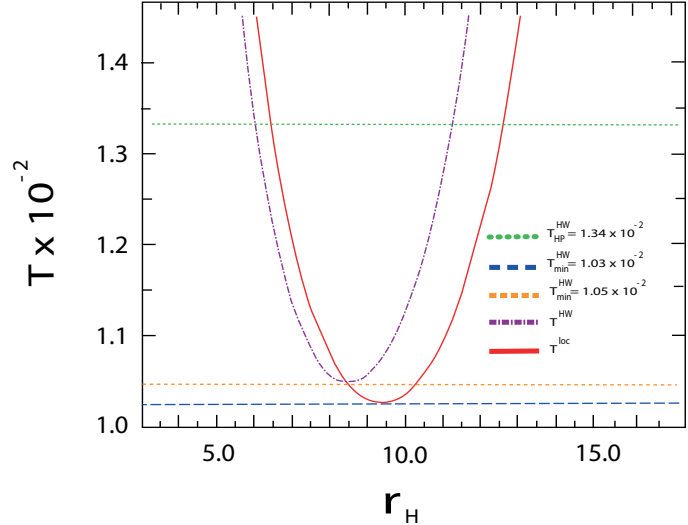}
\caption{Graph for T for  each n=2, $\kappa \eta^2 = 0.3  \cdot 10^{-5}$ and $\psi_n =-0.4 \cdot 10^{-2n}$  with r=105.57.}
 \label{Figure14}
 \end{figure}

So far, the case analyzed considers $ \psi_n <0 $, which can be treated by the Hawking's formalism. In the case of the local formalism, it is possible to verify that independently of $ \psi_n $, it was possible to study both behaviors.

In the graph of Figure 14, the study of the compatibility between the two formalisms can be found. It can be verified that both present the same critical behavior if we choose the value in the local quantities of the fixed point where  local quantities are calculated.

In Figure 14, it can be seen that both the case of $ \psi_n <0 $ to local and Hawking cases, can present the same set of minima and Hawking-Page temperature, showing compatibility between the two theories. It is always possible to find the same set of values if we regulate the values of $r$ in local formalism. In this Figure, this happens when, for a given Hawking-Page temperature, in local formalism, we set r so that the Hawking-Page temperature value is consistent with Hawking's formalism.

  \begin{figure}[htb]
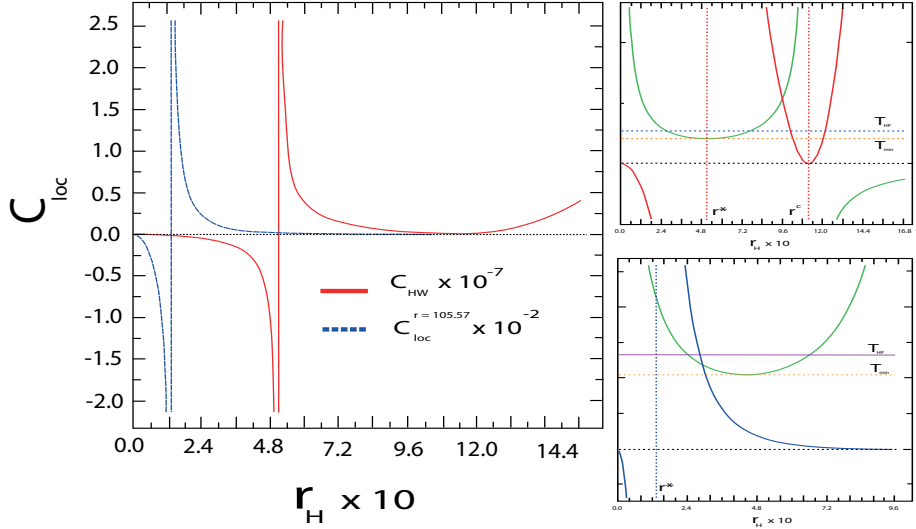

\includegraphics[width=8cm, height=7cm]{fig15a-1.pdf}
\includegraphics[width=4cm, height=7cm]{fig15b-1.pdf}
\caption{Graph for the comparation between  $C_{loc}$ and $C_H$ for  each n=2, $\kappa \eta^2 = 0.3  \cdot 10^{-5}$ and $\psi_n =-0.4 \cdot 10^{-2n}$ with $T_{HP} = 1.3408 \times 10^{-2}$ for both cases. }
 \label{Figure15}
 \end{figure}

In Figure 15, a comparison was made between the heat capacity calculated in Hawking's formalism ($ C_{ HW}$) and Local ($ C_{loc}$) in the case of psi <0. In the figure to the left it can be seen that in the case of Hawking formalism, as indicated in the graph, there are three distinct regions, the first comprises a negative $C_{ HW}$ for $ r_H <r ^ * $. The other region corresponds to a positive $ C_{HW}$ for $ r^* <r _H < r^ c $  and the latter comprises the region given by $ r_H > r^c$. In the figure to the right, upper part, this behavior was further detailed. Analyzing the graph to the right, it can be clearly seen that $ r ^ * $ coincides with the forbidden / allowed transition region. The region $rH <r ^ *$ is a forbidden layer because it has specific  heat negative. The region $r ^* <r_H <r ^ c $ is permissive where we find a stable black hole. In this region the temperature starts from the minimum and grows with the radius of the horizon while the heat capacity decreases.
In this region the black hole is stable. When $ r_H> r ^ c $ we have again a forbidden region.

The other curve of the specific heat of Figure 15 left corresponds to the local case. For the comparison the parameters of the local model were adjusted to  $r = 105.57$. With these conditions, the Hawking-Page temperature can be obtained. In figure 15 the right panel down can analyze the detail of the local case. It can be seen that in this case the solution to the black hole presents two possibilities corresponding to both heat capacity and temperature greater than zero. It can be seen that there is a region between $r * <r_H < r_{min} $, where a black hole solution is possible, but this is unstable because the temperature decreases with the radius of the horizon going to a more stable location that is $r_{min}$. this region is called a small black hole. The other region presents, as in the case of the Hawking formalism, the temperature increasing with the increase of $r_h$, showing that for the increase of the black hole this is giving more energy, for that reason warmer. This region is called the Great Black Hole.

In Figure 16, for n = 2, the comparison of the black hole in the local formalism with $\psi <0$ and $\psi> 0$  for the same Hawking-Page temperature was analyzed. As in the case of figure 15, where the $\psi <0$ case was analyzed, it can be seen that the limits for large $ r_H$ depend on the value of r. Comparing these two cases, it can be analyzed that when $ \psi> 0$, has only one allowed region for $r * <r_H <r^c $, where we have both temperature and calorific capacity positive. In this region there may be several possibilities, the temperature decreases as the heat capacity decreases, for increasing $ r_H$, small black hole. A region where temperature increases and specific heat decreases and a region where both specific heat and temperature increase with increasing horizon radius $r_H$ this region is the big black hole.

 \begin{figure}[htb]
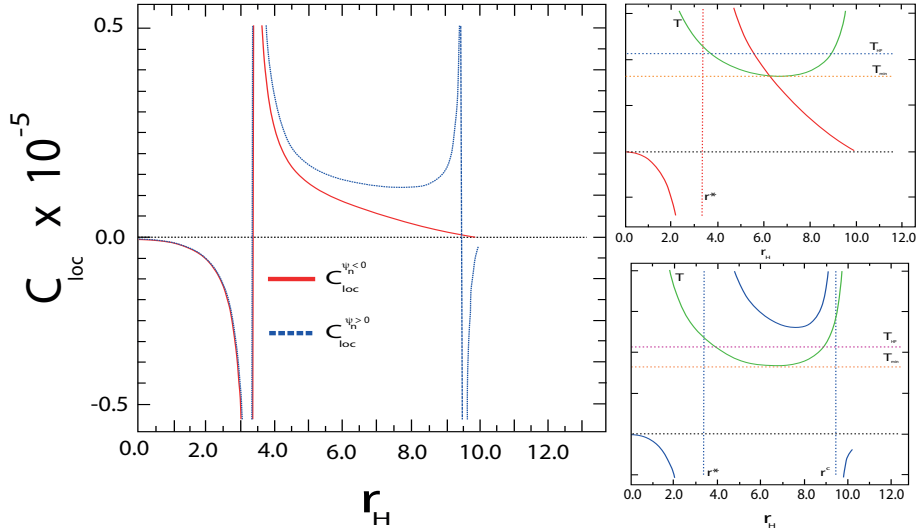

\includegraphics[width=8cm, height=7cm]{fig16a-1.pdf}
\includegraphics[width=4cm, height=7cm]{fig16b-1.pdf}
\caption{Graph for the comparation between  $C_{loc}$ for $\psi_n<0$ and $C_{loc} $ for $ \psi_n>0$  each n=2, $\kappa \eta^2 = 0.3  \cdot 10^{-5}$ and $\psi_n =\mp 0.4 \cdot 10^{-2n}$ r=10 and $T_{HP} = 0.1078$.}
 \label{Figure16}
 \end{figure}

\section{Closing Remarks}

In this work, the solution of a black hole in a region containing a global monopole in a class of f(R) gravity models has been investigated using a power law approach. To pursue our investigation, we have considered the weak field approximation for small radial coordinate values, assuming the validity of the condition of the eq. (\ref{rlimits}). In this his region, it is possible to analyze the behavior of BH solutions for $ n> 0 $ and $ \psi_n $ positive and negative.

The very goal of this work is the analysis of the thermodynamical behavior of the black hole, considering two possible formalisms: Hawking's, on the one hand, and local formulation on the other hand.

In the case of Hawking's approach, it was possible to study the case where $ \psi_n <0 $ as being stable, by studying the positivity of the temperature with the positivity of the heat capacity. This black hole was called a large black hole, located for the radius of the horizon, $ r_H> r_{min} $. For $ r_H <r_{min} $, we have an unstable black hole despite having a region between $ r_H ^ * <r_H <r_{min} $, where we the heat capacity is positive and the temperature is positive.

The case with $ \psi_n> 0 $ has been tackled in the local formalism; this framework is actually valid for both $ \psi_n $ positive and negative. The local case was studied in more details for $ \psi_n> 0$ and a BH with similar characteristics as in the  Hawking's treatment was obtained, if we work in the region of validity. In this work, we also compared these two regimes, by analyzing not only the relationship between heat capacity and temperature, but also the phase transitions that the system may undergo. It has been found that, for small values of $ r_H $, the two formalisms are compatible. When $ r_H $ grows, it reaches a value where the BH undergoes another transition, now related to a change of behavior in the heat capacity. In the case of the thermodynamics with the Hawking formalism, the heat capacity passes from a region where it decreases with increasing temperature to a region where it increases with temperature. In the local case the heat capacity for $ r_H> r_{min}$ always decreases with temperature going to negative values.

Based on the assumption that heat capacity and temperature are not sufficient to guarantee stability, the Hawking-Page phase transition, which applies to both Hawking's formalism and local formalism, was also considered. It has been checked that, in both formalisms, a Hawking-Page phase transition exists as another attribute in favor of the stability of BH.

The case with $ n = 2 $ has also been contemplated. The relevance of this case lies in the fact that the power law parameter, $ \psi_2 $, can act as a cosmological constant, that can be positive or negative in this region.

The idea now is to inspect the asymptotic region that occurs for the case $ n <0 $. With this analysis, it will be possible to study the limits of-Sitter and Anti-de Sitter and the behavior on the borders. Besides the asymptotic behaviors, it is also interesting to study the potential generators of the power law presage that can be clarified when discussing the problem all over the space.
In future works, we intend to investigate this case, that might be important for the AdS/CFT context for the $n=2$ situation. These aspects are under study and shall be the subject of a forthcoming work. It is possible to solve the stability problem for a negative $\psi_2$ and to use the same thermodynamical procedure that we have used here. The importance, in the AdS context, is the fact that this black hole presents an angular deficit, and it is important to study the holographic principle of this object. We have shown in \cite{Bayona:2010sd} that the structure of the angular deficit in a theory in $D = (1 + 3)$ is preserved on the $(1 + 2)$ boundary and, in our case, the defect is preserved in $D = (1 + 2)$.  The AdS4/CFT3 is very important in studying planar materials like graphene and topological insulators. The introduction of the defect in this type of systems gives us current properties in these materials and it may become a relevant aspect of our work, opening up new application possibilities for our model. Another issue that we can pick up to study is the behavior of the objects near these systems \cite{BallonBayona:2013gx} with both regimes for $\psi_2$. We can intensively study quark/anti-quark interactions and discuss the confining and non-confining transitions as well as the chiral symmetry breaking critical temperature \cite{BallonBayona:2008uc}.

 \vspace{1 true cm} { \bf ACKNOWLEDGEMENTS: }
F. B. Lustosa would like to thank CNPq-Brasil for his MSc Fellowship at the Instituto de F\'{\i}sica of Universidade Federal Fluminense. The authors express their gratitude to T.R.P. Caram\^es for clarifying discussions.


\begin{thebibliography}{99}
\footnotesize



\bibitem{Buchdahl:1983zz}
  H.~A.~Buchdahl,
  Mon.\ Not.\ Roy.\ Astron.\ Soc.\  {\bf 150}, 1 (1970).


\bibitem{Morais:2015ooa}
  J.~Morais, M.~Bouhmadi-López and S.~Capozziello,
  JCAP {\bf 1509}, no. 09, 041 (2015)
  [arXiv:1507.02623 [gr-qc]].

\bibitem{Takahashi:2015ifa}
  K.~Takahashi and J.~Yokoyama,
  Phys.\ Rev.\ D {\bf 91}, no. 8, 084060 (2015)
  [arXiv:1503.07412 [gr-qc]].


\bibitem{Nojiri:2010wj}
  S.~Nojiri and S.~D.~Odintsov,
  Phys.\ Rept.\  {\bf 505}, 59 (2011)
  [arXiv:1011.0544 [gr-qc]].

\bibitem{Sotiriou:2008rp}
  T.~P.~Sotiriou and V.~Faraoni,
  Rev.\ Mod.\ Phys.\  {\bf 82}, 451 (2010)
  [arXiv:0805.1726 [gr-qc]].

\bibitem{Perlmutter:1998np}
  S.~Perlmutter {\it et al.}  [Supernova Cosmology Project Collaboration],
  Astrophys.\ J.\  {\bf 517}, 565 (1999)
  [astro-ph/9812133].



\bibitem{Hu:2007nk}
  W.~Hu and I.~Sawicki,
  Phys.\ Rev.\ D {\bf 76}, 064004 (2007)
  [arXiv:0705.1158 [astro-ph]].

\bibitem{Appleby:2007vb}
  S.~A.~Appleby and R.~A.~Battye,
  Phys.\ Lett.\ B {\bf 654}, 7 (2007)
  [arXiv:0705.3199 [astro-ph]].


\bibitem{Starobinsky:2007hu}
  A.~A.~Starobinsky,
  JETP Lett.\  {\bf 86}, 157 (2007)
  [arXiv:0706.2041 [astro-ph]].



\bibitem{Ohta:2015efa}
  N.~Ohta, R.~Percacci and G.~P.~Vacca,
  Phys.\ Rev.\ D {\bf 92}, no. 6, 061501 (2015)
  [arXiv:1507.00968 [hep-th]].


\bibitem{corda} 
C. Corda, Int. J. Mod. Phys. D, {\bf 18}, 2275 (2009) [arXiv:0905.2505[gr-qc]].


\bibitem{Bekenstein:1972tm}
  J.~D.~Bekenstein,
  Lett.\ Nuovo Cim.\  {\bf 4}, 737 (1972).


\bibitem{Bekenstein:1973ur}
  J.~D.~Bekenstein,
  Phys.\ Rev.\ D {\bf 7}, 2333 (1973).


\bibitem{Bekenstein:1974ax}
  J.~D.~Bekenstein,
  Phys.\ Rev.\ D {\bf 9}, 3292 (1974).

\bibitem{Hawking:1974sw}
  S.~W.~Hawking,
  Commun.\ Math.\ Phys.\  {\bf 43}, 199 (1975)
  [Erratum-ibid.\  {\bf 46}, 206 (1976)].


\bibitem{Cai:2009ua}
  R.~-G.~Cai, L.~-M.~Cao and N.~Ohta,
  JHEP {\bf 1004}, 082 (2010)
  [arXiv:0911.4379 [hep-th]].


\bibitem{Myung:2008eb}
  Y.~S.~Myung, Y.~-W.~Kim and Y.~-J.~Park,
  Phys.\ Rev.\ D {\bf 78}, 084002 (2008)
  [arXiv:0805.0187 [gr-qc]].

\bibitem{Biswas:2010zzb}
  R.~Biswas and S.~Chakraborty,
  Gen.\ Rel.\ Grav.\  {\bf 43}, 41 (2011).



\bibitem{delaCruzDombriz:2009et}
  A.~de la Cruz-Dombriz, A.~Dobado and A.~L.~Maroto,
  Phys.\ Rev.\ D {\bf 80}, 124011 (2009)
  [Erratum-ibid.\ D {\bf 83}, 029903 (2011)]
  [arXiv:0907.3872 [gr-qc]].


\bibitem{Kibble:1976sj}
  T.~W.~B.~Kibble,
  J.\ Phys.\ A {\bf 9}, 1387 (1976).


\bibitem{Vilenkin:1984ib}
  A.~Vilenkin,
  Phys.\ Rept.\  {\bf 121}, 263 (1985).

\bibitem{Barriola:1989hx}
  M.~Barriola and A.~Vilenkin,
  Phys.\ Rev.\ Lett.\  {\bf 63}, 341 (1989).


\bibitem{Carames:2011uu}
  T.~R.~P.~Caram\^es, E.~R.~Bezerra de Mello and M.~E.~X.~Guimar\~aes,
  Int.\ J.\ Mod.\ Phys.\ Conf.\ Ser.\  {\bf 3}, 446 (2011)
  [arXiv:1106.4033 [gr-qc]].


\bibitem{Marunovic:2013eka}
  A.~Marunovic and M.~Murkovic,
  Class.\ Quant.\ Grav.\  {\bf 31}, 045010 (2014)
  [arXiv:1308.6489 [gr-qc]].




\bibitem{Multamaki:2006zb}
  T.~Multamaki and I.~Vilja,
  Phys.\ Rev.\ D {\bf 74}, 064022 (2006)
  [astro-ph/0606373].





\bibitem{Carames:2011xi}
  T.~R.~P.~Caram\^es, E.~R.~Bezerra de Mello and M.~E.~X.~Guimar\~aes,
  Mod.\ Phys.\ Lett.\ A {\bf 27}, 1250177 (2012)
  [arXiv:1111.1856 [gr-qc]].





\bibitem{Yu:1994fy}
  H.~-W.~Yu,
  Nucl.\ Phys.\ B {\bf 430}, 427 (1994).



\bibitem{Faraoni:2010yi}
  V.~Faraoni,
  Entropy {\bf 12}, 1246 (2010)
  doi:10.3390/e12051246
  [arXiv:1005.2327 [gr-qc]].



\bibitem{Davies78}  P. C. W. Davies, Rep. Prog. Phys., Vol. 41, (1978).

\bibitem{Bardeem73} J. M. Bardeen;B. Carter; S. W. Hawking, Communications in Mathematical Physics 31, (1973), no. 2, 161?170.


\bibitem{Hawking74} S. W. Hawking, Nature 248, 30 - 31, (1974).




\bibitem{Man:2013sf}
  J.~Man and H.~Cheng,
  Phys.\ Rev.\ D {\bf 87}, no. 4, 044002 (2013)
  [arXiv:1301.2739 [hep-th]].



\bibitem{Tolman:1930zza}
  R.~C.~Tolman,
  Phys.\ Rev.\  {\bf 35}, 904 (1930).

\bibitem{Bayona:2010sd}
  C.~A.~B.~Bayona, C.~N.~Ferreira and V.~J.~V.~Otoya,
  Class.\ Quant.\ Grav.\  {\bf 28}, 015011 (2011)
  doi:10.1088/0264-9381/28/1/015011
  [arXiv:1003.5396 [hep-th]].

\bibitem{BallonBayona:2013gx}
  A.~Ballon-Bayona, C.~N.~Ferreira and V.~J.~V.~Otoya,
  Phys.\ Rev.\ D {\bf 87}, no. 10, 106007 (2013)
  doi:10.1103/PhysRevD.87.106007
  [arXiv:1302.0802 [hep-th]].

\bibitem{BallonBayona:2008uc}
  C.~A.~Ballon Bayona and C.~N.~Ferreira,
  Phys.\ Rev.\ D {\bf 78}, 026004 (2008)
  doi:10.1103/PhysRevD.78.026004
  [arXiv:0801.0305 [hep-th]].



\end{thebibliography}
\end{document}